\documentclass{elsart1p}

\usepackage{graphicx}
\usepackage{epsfig}
\usepackage{subfigure}
\usepackage{amssymb}

\begin{document}

\begin{frontmatter}

\title{On the nucleation of hadronic domains \\ in the quark-hadron transition}

\author[UFRJ]{B. W. Mintz}
\author[USP]{$\!\!$, A. Bessa}
\author[UFRJ]{$\!\!$, E. S. Fraga}

\address[UFRJ]{Instituto de F\'\i sica, Universidade Federal do Rio de Janeiro, \\
Caixa Postal 68528, Rio de Janeiro, RJ, 21941-972, Brazil}
\address[USP]{Instituto de F\'{\i}sica, Universidade de S\~ao Paulo\\
Caixa Postal 66318, S\~ao Paulo, SP, 05315-970, Brazil}

\begin{abstract}
 We present numerical results on bubble profiles, nucleation rates and time evolution for 
a weakly first-order quark-hadron phase transition in different expansion scenarios. We 
confirm the standard picture of a cosmological first-order phase transition, in which 
the phase transition is entirely dominated by nucleation. We also show that, even for expansion 
rates much lower than those expected in heavy-ion collisions nucleation is very unlikely, indicating
 that the main phase conversion mechanism is spinodal decomposition.
\end{abstract}

\end{frontmatter}

\section{Introduction}

Although Lattice QCD results seem to indicate a crossover for the quark-hadron transition 
\cite{Karsch:2008fe}, the possibility of a weakly first-order transition is not ruled 
out yet from the experimental point of view and some observables behave differently 
for a first-order transition and for a crossover \cite{Huovinen2005}. 
Moreover, Lattice QCD does not provide any information on the (nonequilibrium) 
dynamical nature of the transition, whose critical behavior could be very 
different from what one might expect from a crossover in the (equilibrium) phase 
diagram.

We calculate exact critical bubble profiles and their critical radii and surface 
tension as functions of the temperature \cite{Bessa_appear} using realistic equations 
of state (EoS). We also compute the fraction of the plasma hadronized via nucleation 
and its temperature as functions of time in order to analize quantitatively the importance 
of nucleation in the dynamics of the phase transition.

\section{A model for the phase transition of a QGP in expansion}

After a hot plasma of quarks and gluons (QGP) is formed, it cools down due to expansion 
and eventually reaches the critical temperature $T_c$. From this point on, the plasma supercools 
and, as it becomes metastable, nucleation starts to act effectively. When one critical 
bubble is actually formed and the corresponding region of the plasma is hadronized, the 
latent heat released in the process may reheat the plasma. This goes on until the 
whole plasma is converted to the hadronic phase and then the temperature suddenly falls. However, 
if the expansion is fast enough, the released heat cannot keep the system from supercooling down 
to some temperature under which it becomes unstable and suffers spinodal decomposition. 
In this case, usually called a quench, nucleation is a secondary mechanism in the phase 
conversion \cite{Mishustin1999,Scavenius_etc2001}.

Our criterion for determining the importance of nucleation relies on the fraction $f(t)$ 
of the plasma which is hadronized from the moment $T=T_c$ and before the unstable spinodal 
temperature is reached.

We assume local thermal equilibrium and (approximate) entropy conservation. For the 
quark phase ($N_f=2+1$), we use two different equations of state: one corresponding to 
lattice results \cite{Cheng:2007jq} and the other one to a Bag Model \cite{CsernaiKapusta1992}. 
For the hadronic phase, we take a gas of over 250 free massive resonances \cite{Kodama_pc}. 
We use the 3-d Euclidean action \cite{Linde1983}
\begin{equation}\label{eq:action}
 S_3[\phi] = \int d^3r\left[\frac12[\nabla \phi({\bf r})]^2 + V[\phi({\bf r}),T]\right] ,
\end{equation}
where $V$ is a phenomenological effective potential given by
\begin{equation}
 V(\phi,T) = a(T)\phi^2 - bT\phi^3 + c \phi^4 .
\end{equation}
We assume that the quark phase corresponds to the order parameter $\phi_q(T) = 0$ and the 
hadron phase to $\phi_h(T)$. The connection with thermodynamics is made through 
the identification $p_q(T)-p_j(T)\equiv V(\phi_j, T)$ for a homogeneous phase with $\phi=\phi_j$. 

We also suppose a homogeneous and isotropic (Hubble-like) expansion of the plasma, with 
scale factor $a(t) = a_0\exp(H_0 t)$, where $H_0$ is a constant. A final assumption 
is the homogeneous reheating of the expanding plasma due to the release of latent heat. All 
these ingredients together lead to the equation \cite{Megevand2008}
\begin{equation}\label{eq:temperature}
 T(t) = \left[\left(T_c\frac{a(0)}{a(t)}\right)^3 + f(t)\frac{\Delta s(T(t))}{s(T(t))}T^3(t) \right]^{1/3},
\end{equation}
where $s=(1-f)s_q+fs_h$ is the spatial average of the entropy density, $s_q(T)$ ($s_h(T)$) is 
the entropy density of the quark (hadron) phase and $\Delta s(T) = s_q(T)-s_h(T)$.


\section{Method for calculating T(t)}
\label{sec:T(t)}

\subsection{Bubble features}
\label{sub:bubble}
Critical bubbles are spherically symmetric extrema of the action (\ref{eq:action}), so that
\begin{equation}\label{eq:eq_dif}
\frac{d^2\phi}{dr^2} + \frac2r\frac{d\phi}{dr}  = \frac{\partial V}{\partial \phi},\;\;\;\;{\rm with}\;\;\;\;
\phi(0) = \phi_h\;\;\;\;\; {\rm and} \;\;\;\;\; \phi(\infty) = \phi_q = 0.
\end{equation}

This equation may be solved analytically for temperatures close to $T_c$ in the 
so-called thin-wall approximation \cite{CsernaiKapusta1992, Linde1983}. However, 
this approximation, which is valid for $R_c/\xi \gg 1$ ($\xi\sim T^{-1}$), fails 
for temperatures much lower than 
$T_c$ and one has to solve eq. (\ref{eq:eq_dif}) numerically. This is what we do in order to 
calculate the main ingredients for the dynamics of the phase conversion: 
the critical bubble radius $R_c(T)$ and surface tension $\sigma(T)$.

\subsection{Dynamical quantities}

In order to calculate $f(t)$, a necessary ingredient for $T(t)$ according to eq. (\ref{eq:temperature}), 
we use \cite{GuthWeinberg1981}
\begin{equation}\label{eq:f(t)}
 f(t) = 1 - \exp\left\{-\int_0^t dt'\, \left(\frac{a(t')}{a(t)}\right)^3 \Gamma[T(t')] \frac{4\pi}{3} R^3(t',t) \right\},
\end{equation}
where the nucleation rate per unit volume per unit time is \cite{CsernaiKapusta1992}
\begin{equation}\label{eq:Gamma}
 \Gamma(T) = \frac{16}{3\pi}\left(\frac{\sigma}{3T}\right)^{3/2}\frac{\sigma\eta R_c}{\xi^4(\Delta\omega)^2} e^{-S_3/T} ,
\end{equation}
and the radius of a bubble that was “born” at time $t'$ with radius $R_0[T(t')]$ and grew 
until time $t$ with velocity $v_w(T) = -V(\phi_h,T)/[\sigma(T) T/2]$ is \cite{Megevand2008}
\begin{equation}\label{eq:R_t}
 R(t',t) = \frac{a(t)}{a(t')}R_0[T(t')] + \int_{t'}^{t}dt''\,v_w[T(t'')]\frac{a(t)}{a(t'')}.
\end{equation}

\section{Results and discussion}
We have solved numerically the set of equations (\ref{eq:temperature}) - (\ref{eq:R_t}) for 
the temperature as a function of time: $T(t)$. 


\begin{figure}[htbp]
  \begin{minipage}[b]{0.47\linewidth}
    \includegraphics[width=0.87\linewidth,angle=270]{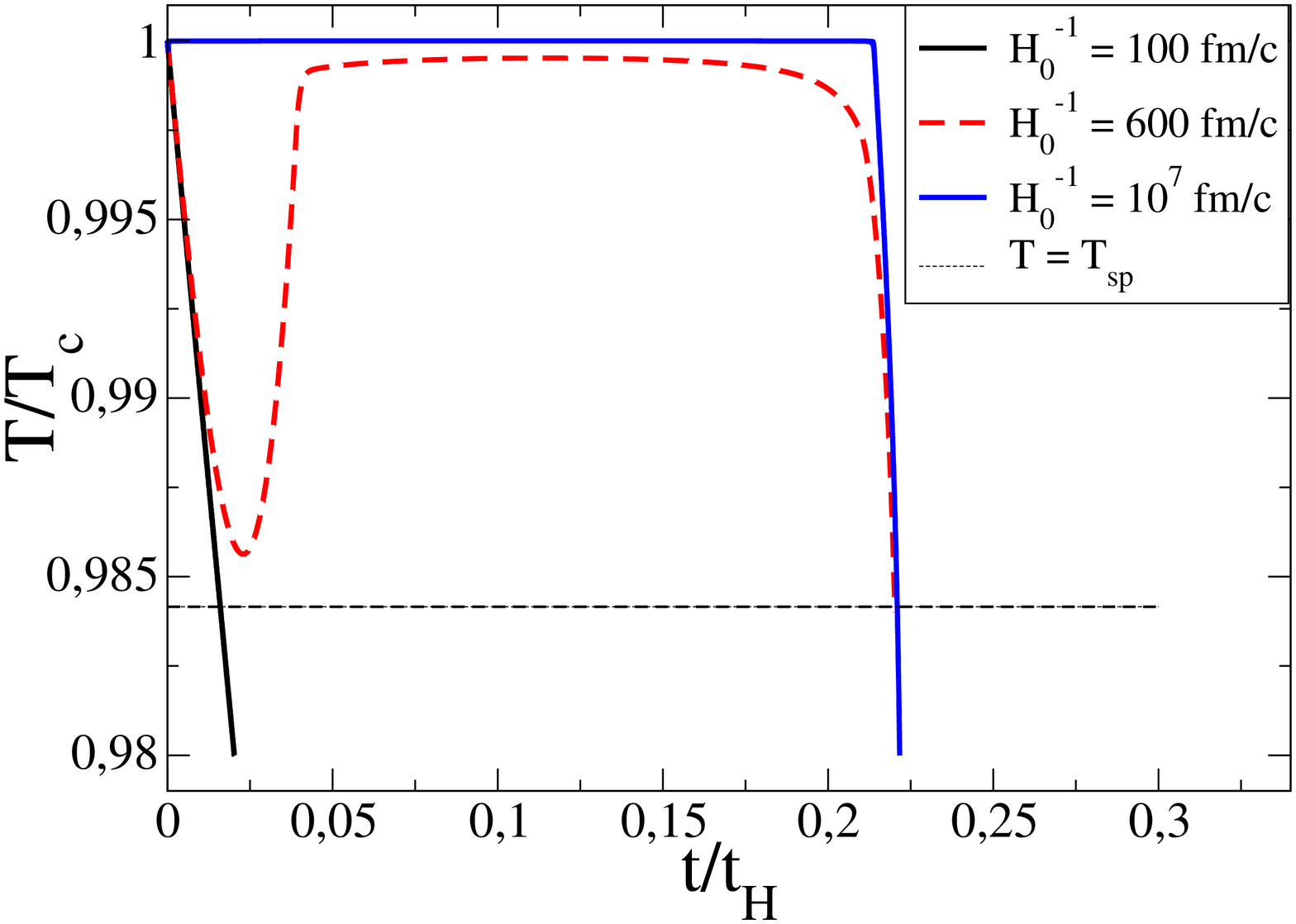}
    \caption{$T(t)$ for different values of the expansion rate $H_0$.}
    \label{fig:Temp_T_mixed}
  \end{minipage}
  \hfill
  \begin{minipage}[b]{0.47\linewidth}
    \includegraphics[width=0.87\linewidth,angle=270]{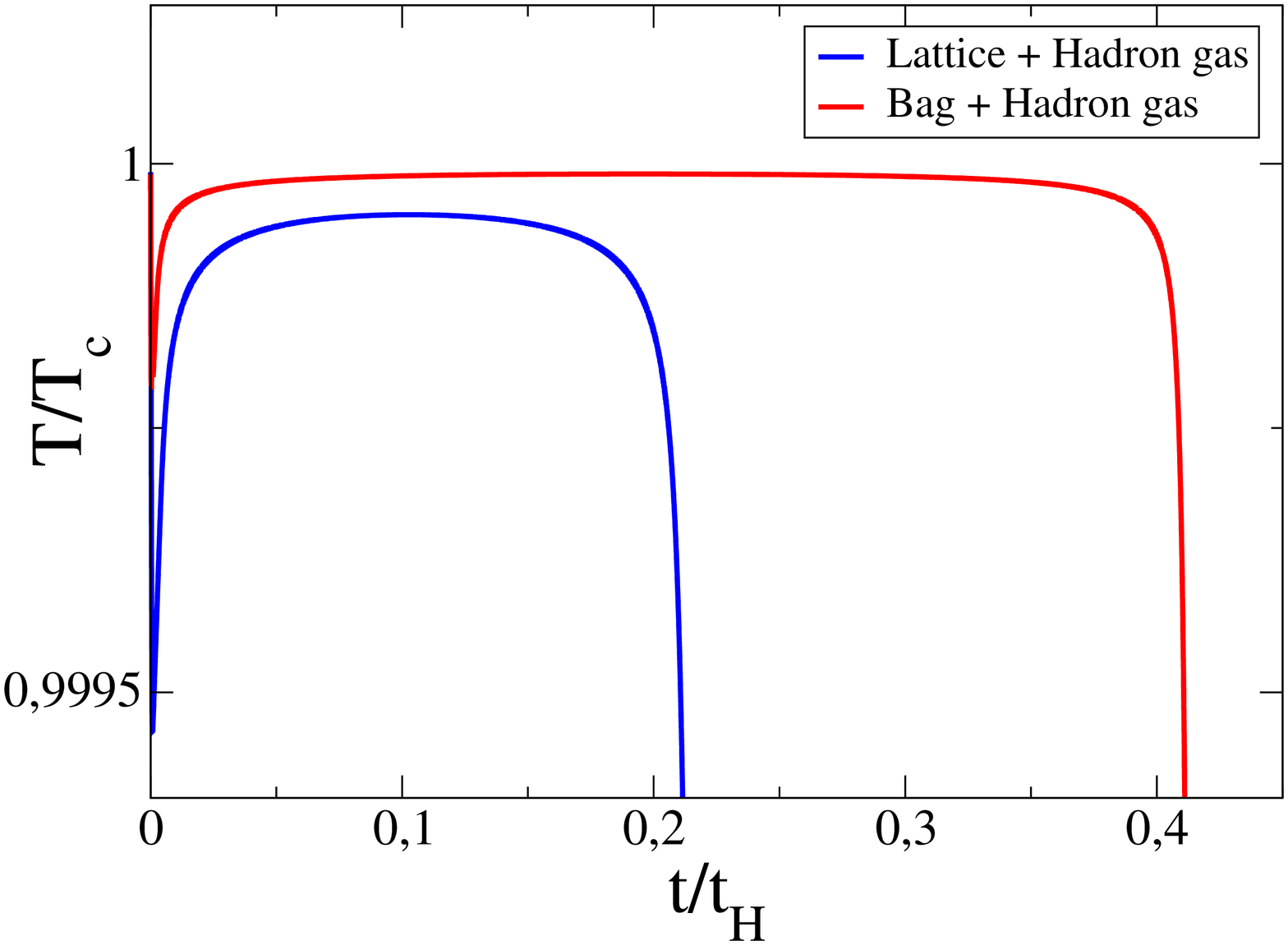}
    \caption{$T(t)$ for different equations of state ($H_0^{-1}=600$fm/c).}
    \label{fig:Temp_T_EoS}
  \end{minipage}
\end{figure}

In Fig. \ref{fig:Temp_T_mixed}, $T(t)$ is shown for three different expansion rates: very fast 
($H_0^{-1}=100\,$fm/c), intermediate ($H_0^{-1}=600\,$fm/c) and very slow ($H_0^{-1}=10^7$fm/c). 
Notice that the plasma suffers a quench into the spinodal region in the fast expansion, for 
it does not have enough time either to nucleate a non negligible number of bubbles or to reheat. 
In the very slow case, the system presents a very slight supercooling which is quickly washed 
away by the plasma reheating, so that the phase conversion follows quite closely the equilibrium, 
as expected from a very slowly expanding system. Inverse expansion rates of the order of $600$fm/c 
(for the Hadron+Lattice, or HL, EoS) lead to the most interesting scenarios, in which the plasma reaches a 
considerable supercooling but reheats before reaching the spinodal temperature. After this reheating, 
the temperature once again approaches $T_c$ and the phase conversion is completed mainly by the  percolation of the bubbles born in the supercooling stage. 

In Fig. \ref{fig:Temp_T_EoS} we show the effect of the latent heat on the time evolution 
of the temperature. The HL EoS leads to a stronger supercooling and a shorter 
coexistence time in comparison with the Hadron+Bag (HB) EoS, which has a larger latent heat. Once 
the HL EoS leads to a smaller latent heat, it takes longer to reheat the plasma, 
and it spends more time in a low temperature (and high nucleation rate) regime. This means that 
when the system is successfully reheated, a large fraction of it is already hadronized and the 
remaining fraction is not able to maintain the expanding system with $T\lesssim T_c$ for much time. 
On the other hand, when the latent heat is larger (e.g., with the HB EoS), the plasma 
reheats sooner and can keep its temperature high for a longer time.

\section{Conclusion}
Using realistic equations of state (in comparison with the usual Bag Model EoS), we showed 
that if QCD has a weakly first-order phase transition, then nucleation is very unlikely the 
main phase conversion mechanism in HIC experiments, regardless of reheating effects. In this 
scenario, the adequate description for the transition should be a quench followed by spinodal 
decomposition. In the early Universe, however, we showed that nucleation completely dominates 
the conversion dynamics. Finally, we showed that a stronger transition leads to a smaller 
supercooling and to a larger coexistence time. 

{\it Acknowledgements}
We would like to thank Takeshi Kodama for discussions and the Brazilian agencies CAPES, CNPq, 
FAPERJ, FAPESP and FUJB for financial support.

\end{document}